\documentclass[aps,prb,twocolumn,superscriptaddress]{revtex4-2}
\usepackage{graphicx}
\usepackage[margin=1in]{geometry}
\usepackage{sidecap}
\usepackage[capbesidesep=quad]{floatrow}
\usepackage{setspace}
\usepackage[colorlinks=true,allcolors=blue]{hyperref}
\usepackage{lineno}
\usepackage{comment}
\usepackage{titlesec}
\titlespacing*{\section}{0pt}{\baselineskip}{\baselineskip}
\titlespacing*{\subsection}{0pt}{\baselineskip}{\baselineskip}

\newcommand{\cusulf}{CuSO$_4\!\cdot\!5$H$_2$O}
\newcommand{\magcro}{MgCr$_2$O$_4$}

\begin{document}
\title{Cryogenic platform to investigate strong microwave cavity-spin coupling in correlated magnetic materials}
\author{Aulden K. Jones}
\affiliation{School of Physics, Georgia Institute of Technology, Atlanta, GA 30332, USA}
\affiliation{Center for Integrated Nanotechnologies, Sandia National Laboratories, Albuquerque, New Mexico 87185, USA}
\author{Martin Mourigal}
\affiliation{School of Physics, Georgia Institute of Technology, Atlanta, GA 30332, USA}
\author{Andrew M. Mounce}
\affiliation{Center for Integrated Nanotechnologies, Sandia National Laboratories, Albuquerque, New Mexico 87185, USA}
\author{Michael P. Lilly}
\affiliation{Center for Integrated Nanotechnologies, Sandia National Laboratories, Albuquerque, New Mexico 87185, USA}
\email{auldenjones@gatech.edu}

\begin{abstract}
We present a comprehensive exploration of loop-gap resonators (LGRs) for electron spin resonance (ESR) studies, enabling investigations into the hybridization of solid-state magnetic materials with microwave polariton modes. The experimental setup, implemented in a \textit{Physical Property Measurement System} by Quantum Design, allows for ESR spectra at temperatures as low as 2 Kelvin. The versatility of continuous wave ESR spectroscopy is demonstrated through experiments on \cusulf\ and \magcro, showcasing the g-tensor and magnetic susceptibilities of these materials. The study delves into the challenges of fitting spectra under strong hybridization conditions and underscores the significance of proper calibration and stabilization. The detailed guide provided serves as a valuable resource for laboratories interested in exploring hybrid quantum systems through microwave resonators.
\end{abstract}


\maketitle


\section{Introduction} 
Electron spin resonance (ESR) spectroscopy is a versatile experimental technique to study and control the dynamics of spin-dense and dilute material systems through microwave-induced spin transitions under an applied magnetic field. The technique can be implemented in continuous wave (CW) and pulsed forms, providing direct determinations of a system's $g$-tensor, relaxation times, and magnetic susceptibilities. The low-frequency nature of ESR in the X-band ($\approx 10~{\rm GHz}$), requires microwave resonators. This can be seen through the relative frequency ($\nu_0$) of the induced $\Delta m_s=\pm1$ transitions compared to the thermal energy of the samples at equilibrium ($1\:{\rm GHz}\!\approx\!0.048 {\rm K}$). In the $k_{\rm B}T\!\gg\!\hbar\nu_0$ regime, thermal excitations lead to weak population differences and absorption between magnetic states \cite{CW_ESR_Est}. By enhancing light-matter interactions, resonators produce a corresponding increase in microwave sensitivity, allowing small changes in sample inductance to be detected. 

Recent experimental work has focused on studying magnetic quantum phases and hybrid quantum states that emerge from spin-microwave photon interactions in cavities. Applications range from quantum information and material control~\cite{ball_loop-gap_2018,everts_ultrastrong_2020} to understanding exotic magnetism \cite{LiberskyDirectSoft2021,yang_investigation_2023}. But investigating the low-energy spectral response of correlated materials with electron-spin-based resonance techniques has long offered unique advantages, with experiments testing exquisite theories of magnon fractionalization into spinon pairs in low-dimensional spin systems \cite{Oshikawa_AFMspinChain_1999, povarov_modes_2011} and spin-orbital quantum magnets \cite{luo_spinon_2018,Sichelschmidt_2019_NaYbS2}. Sensitivity to the spinon continuum is associated with an increasing resonance linewidth with decreasing temperature. However, in practice, thermal fluctuations obscure the observation of these collective quantum phenomena. Gaining a more precise picture thus requires performing measurements at sub-Kelvin temperatures beyond the limit of most commercial and very high magnetic-field experimental set-ups.

Additionally, strong spin-lattice coupled materials typically show no visible ESR spectra due to their rapid relaxation rates. In these cases, phonons dissipate absorbed microwave energy, resulting in broad spectral peaks that only become observable in the low-temperature, single-Kelvin regime, where relaxation rates slow and lattice vibrations die out \cite{everts_ultrastrong_2020}. This is the case with many rare-earth-based quantum magnets.

This manuscript reports a straightforward cryogenic implementation of loop-gap resonators (LGRs) to investigate the feasibility of different magnetic quantum materials to hybridize with a microwave polariton mode down to $T\approx10$~K. Our work integrates past results demonstrating the utility of LGRs to control hybrid quantum systems~\cite{ball_loop-gap_2018} and to obtain traditional ESR spectra on molecular magnets using a commercial Quantum Design {Physical Property Measurement System (PPMS)}~\cite{joshi_adjustable_2020}. The flexibility of LGRs and our low-cost setup not only enables sub-X-band ESR spectroscopy ($\nu \approx 3$--$6$ GHz) inside a cryostat environment but also serves as a platform to characterize the potential for strong coupling between microwave resonators and spin-dense, correlated materials. We demonstrate our approach on two magnetic insulators described by a Heisenberg exchange Hamiltonian and hosting quantum and classical spin-liquid correlations, respectively. We also review the relevant technical concepts in the hope of helping successful implementations by other laboratories and paving the way for our future implementation in a dilution refrigerator reaching sub-Kelvin temperatures.

    \begin{figure*}[t]
        \centering
        \fbox{\includegraphics[width=.76\columnwidth]{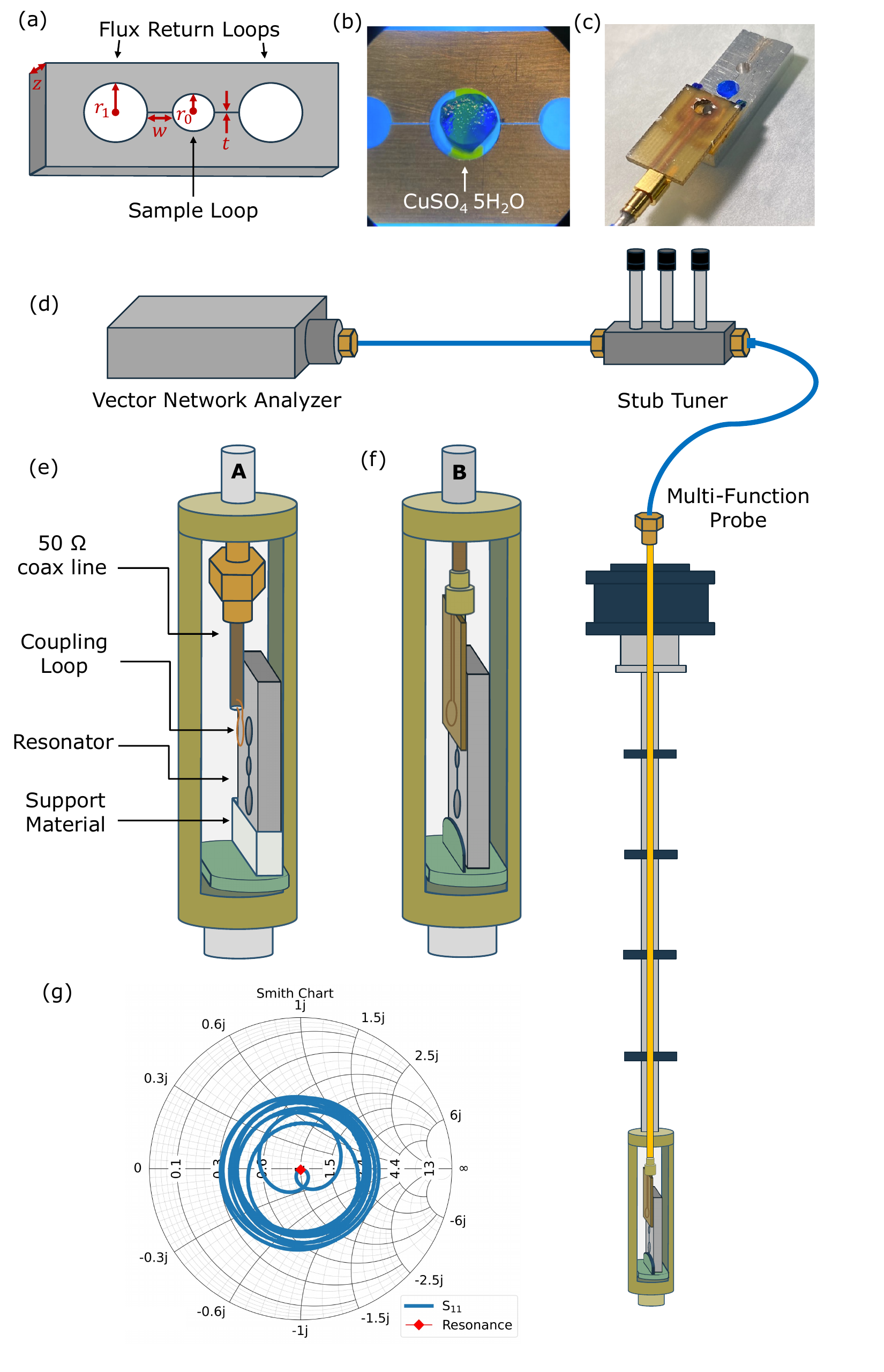}}
        \caption{(Continued on the following page.)}
        \label{fig:FirstFig}
    \end{figure*}
\setcounter{figure}{0}
    \begin{figure*}[t]
        \caption{(a) Loop gap resonator diagram showing the relevant geometric parameters that were selected to obtain a desired resonance frequency as originally described \cite{wood_loop-gap_1984}. The geometric parameters of the resonator used throughout this study are $r_0=2.3$~mm, $r_1=1.5$~mm, $t\approx0.07$~mm, $w=2.5$~mm, $w=4.0$~mm. (b) View of the \cusulf\ sample studied inside the LGR. The sample is secured at the bottom with a strip of kapton tape. The crystal shown here has been cut to form a cylinder extending along the entire $z$ length of the LGR. While no attempt was made to accurately align the triclinic unit-cell for either of the two distinct Cu ions in the crystal structure, it should be noted that the cylindrical axis of the cut crystal falls along the $b$-axis of the triclinic cell~\cite{bissey_differentiated_1995}. Approximating the cavity filling factor by the sample to sample-space volume ratio yields $\eta \approx 0.5$. (c) Image of the \cusulf\ crysral loaded LGR with PCB coupling loop attached outside of the probe. (d) Diagram showing every microwave component utilized in setup B for low-temperature measurements down to $T=10$~K (for data reported here) or $T=1.8$~K (base temperature). The one-port vector network analyzer (VNA) generates and detects the signal, with a maximum frequency up to $\nu_{\rm max}=6$~GHz. The stub-tuner changes the characteristic impedance of the coax transmission line to critically couple to the LGR-sample system at the resonance frequency. This process ensures that the maximum power is transferred to the load, or resonator. Next, the signal is transferred to the sample via a 1.2-meter-long SMA to SMP semi-rigid cable, where it interfaces with the PCB coupling loop. (e) The schematic shows the first experimental setup (setup A) at the bottom of the PPMS multi-function probe (MFP). Arrows label each primary component. Going from top to bottom, the $\mathrm{50 \: \Omega}$ semi-rigid coax fed through the center rod is connected to an additional length of coax via an SMP to SMA adapter. This added section is terminated in a single turn of the center conductor soldered to the outer casing. It forms a loop that transmits the VNA signal to the resonator through its mutual inductance with one of the outer return flux loops. The aluminum loop-gap resonator is secured to the platform by a 3D-printed support box that tightly fits the bottom section of the resonator. This box is attached to the bottom MFP puck with a quick-drying adhesive. (f) Diagram of setup B, with the main improvement being the stabilized PCB coupling loop suspended 1.2 mm above the LGR body via a u-shaped FR4 spacer to prevent electrical contact. Each copper trace on the PCB is terminated onto an SMP male connector that leads to the transmission line. The LGR, spacer, and PCB are all adhered to each other via GE varnish, which provides stability during measurements. The bottom of the LGR is mounted to a right-angled puck attached via GE varnish. (g) Smith chart from the VNA once the experiment is loaded into the cryostat. The blue $\mathrm{S_{11}}$ trace shows how the impedance of the setup changes periodically as a function of frequency, with only the LGR resonance frequency at critical coupling as labeled by the red marker in the center of the plot. The periodicity of the frequency change exactly matched what we would expect for an open coax cable of the length used in these experiments. This particular Smith chart was taken for \magcro\ at room temperature, but a similar form is seen throughout all measurements across temperatures.}
    \end{figure*}

\section{Design Experimental Set-Up and Method}

\subsection{Resonators}

A critical component in conventional ESR spectroscopy and many hybrid quantum systems setups is the microwave resonant structure that hosts the sample or system of interest. Maximizing microwave sensitivity in ESR typically requires maximizing the product of the resonator quality factor ($Q$) and filling factor ($\eta$). For this study, we utilized lumped-element resonators known as loop-gap resonators (LGRs). These structures were first developed in the 1980's for nuclear \cite{Hardy1981SplitRing} and electron \cite{FRONCISZ_1982} magnetic resonance studies in sub X-band frequencies. While LGRs are limited to an accessible frequency range of $\nu_0 = 2$--$8$~GHz and operate with much smaller quality factors ($Q=\nu_0/\Delta\nu$), their significant filling factors ($\eta$) and homogeneous field ($B_{\rm MW}$), makes them ideal for strong coupling to bulk magnetic materials in low-temperature environments. 

The specific design we adopted uses a three-loop, two-gap structure first described in 1984 for ESR and electron-nuclear double resonance spectroscopy (ENDOR) spectroscopy \cite{wood_loop-gap_1984}, and recently utilized in experiments investigating quantum transduction phenomena \cite{ball_loop-gap_2018,everts_ultrastrong_2020}. One of the primary benefits of a multi-loop and gap resonator is the ability to manage and contain the returning magnetic flux pathways in a microwave experiment, making them ideal to use in cryostats and dilution refrigerators where large superconducting magnets limit the free space available to construct an experiment. 

The geometric parameters that determine an empty LGR resonance frequency and quality factor are shown in Fig.~\ref{fig:FirstFig}(a) following their previous description in Ref.~\cite{wood_loop-gap_1984}. For the experiments carried out in this work, aluminum LGRs were created with a manual milling machine in two halves to create $t = 70~\mu{\rm m}$ sized gaps. The two halves were then attached with a silver-based conductive epoxy to form the full resonator. While this manufacturing approach does yield lower quality factors than precision-machined LGRs \cite{ball_loop-gap_2018}, we find them adequate to observe the necessary spectra for our demonstrational experiments; future studies will be carried out on higher-quality resonators. 

While modeling a desired LGR sample size and frequency is possible with the design equations in the early literature \cite{wood_loop-gap_1984}, this only resulted in predictions accurate within $|\nu_{\rm 0}\!-\!\nu_{\rm calc}| \approx$ 1 GHz of the experimental resonance frequency. Instead, we utilized COMSOL Multiphysics to simulate and find the eigen-frequencies and modes for many of our resonator designs. In the following, we refer to $\nu_{\rm 0}$ as the main resonance frequency of interest for our experiments. 

\subsection{Sample Selection}

To explore strong cavity-material coupling in samples that are known to host correlated spin physics, we chose to investigate two transition-metal based Heisenberg antiferromagnets: the one-dimensional $S=1/2$~quantum liquid \cusulf\ and the three-dimensional $S=3/2$~classical spin-liquid \magcro. While these materials have been well studied using a host of experimental techniques, from traditional and high-field ESR to inelastic neutron scattering experiments \cite{mourigal_fractional_2013, bai_magnetic_2019}, our experiments investigate the ability of these systems to strongly hybridize with a cavity mode over a range of temperatures.  These samples were chosen due to their well-characterized ESR response, availability as high-quality large single-crystals fitting our LGR (Fig.~\ref{fig:FirstFig}(b)), and distinctly different magnetic behavior regarding magnetic excitation bandwidth and the role of quantum correlations. 

\subsection{Microwave Generation, Detection and Transport in Cryogenic Environment}

A \textit{R60 1-Port 6 GHz} Vector Network Analyzer (VNA) from Copper Mountain Technologies was used as the microwave source and detector. A diagram of the finalized microwave setup used is shown in Fig.~\ref{fig:FirstFig}(d). The signal from the VNA was fed into a stub tuner using a procedure detailed in Ref.~\cite{Eisenach_2018}. A semi-rigid coax cable was then used to transmit the microwave signal through a hermetically sealed union to the bottom of a Quantum Design cryogen-free \textit{Physical Property Measurement System} (PPMS) equipped with a 14~T superconducting magnet and reaching base temperature of $T\!=\!2$~K. The experiment itself used Quantum Design's \textit{Multi-Function Probe} (MFP), which provided enough space for the sample-LGR assembly. The coupling structure and the right-angle mount secured the LGR with the loops applied perpendicular to the DC field. 

\subsection{Microwave Coupling}

Mutual inductive coupling was utilized between the LGR body and the coaxial transmission line, see e.g. Fig.~\ref{fig:FirstFig}(c). In practice, two different coupling setups were used, labeled setup A and B in Fig.~\ref{fig:FirstFig}(e) and (f), respectively. Maximizing the microwave power transferred to the resonator is important and characterized by the degree to which critical coupling is achieved. Critical coupling, or matching, occurs when the impedance of the transmission line equals that of the resonator. This condition minimizes reflected microwave power, and various methods to create these matching networks for LGRs are well documented \cite{Rinard1993MicrowaveCS}. The coupling for each setup was optimized by analyzing the Smith chart for the VNA frequency range of interest. In setup A, we attempted to critically couple by physically moving the inductive loop on the end of the coax line on the bench top. While this allowed us to get a detectable signal, it was distinctively over-coupled and limited our ability to fine-tune the setup inside the cryostat. We improved flexibility using a stub tuner, which controlled the impedance of the coax line to match the resonator at its resonance frequency, as seen in figure \ref{fig:FirstFig}(d). This facilitated in-situ tuning before the ESR field sweeps inside the cryostat. In setup B, a printed circuit board (PCB) coupling loop was used with an additional FR4 material as a spacer between the LGR body and the copper trace. This provided a more stable configuration that experienced minimal drift as the experiment was cooled from room temperature inside the cryostat. More details of our set-up are provided in the caption of Fig.~\ref{fig:FirstFig}.

\subsection{Sample Preparation and Orientation}

For high sensitivity of ESR response, including at room temperature, the resonators and the samples had to match in size to ensure a high filling factor. For \cusulf, this entailed polishing a large crystal down to the size of the LGR sample loop, while floating-zone grown \magcro\ crystals~\cite{mgcro_crystalgrowth} were observed to already fit well within the first LGR sample radius of 2.3 mm, Fig.~\ref{fig:FirstFig}(b). While \cusulf\ fit well within the resonator, the exact orientation of the crystal was not determined, as an exact $g$-factor determination was second to observing an initial ESR response for the experimental setup. The sample in setup A was adhered to the resonator with Kapton tape, which crossed over one side of the sample loop, and can be seen behind the \cusulf\ crystal in Fig.~\ref{fig:FirstFig}(b). Setup B utilized GE varnish to cement both samples. For \magcro, the nearly Heisenberg nature of the magnetism meant that precise sample alignment was also not necessary for this demonstration study.

    \begin{figure*}[t!]
        \centering
        \includegraphics[width=1\columnwidth]{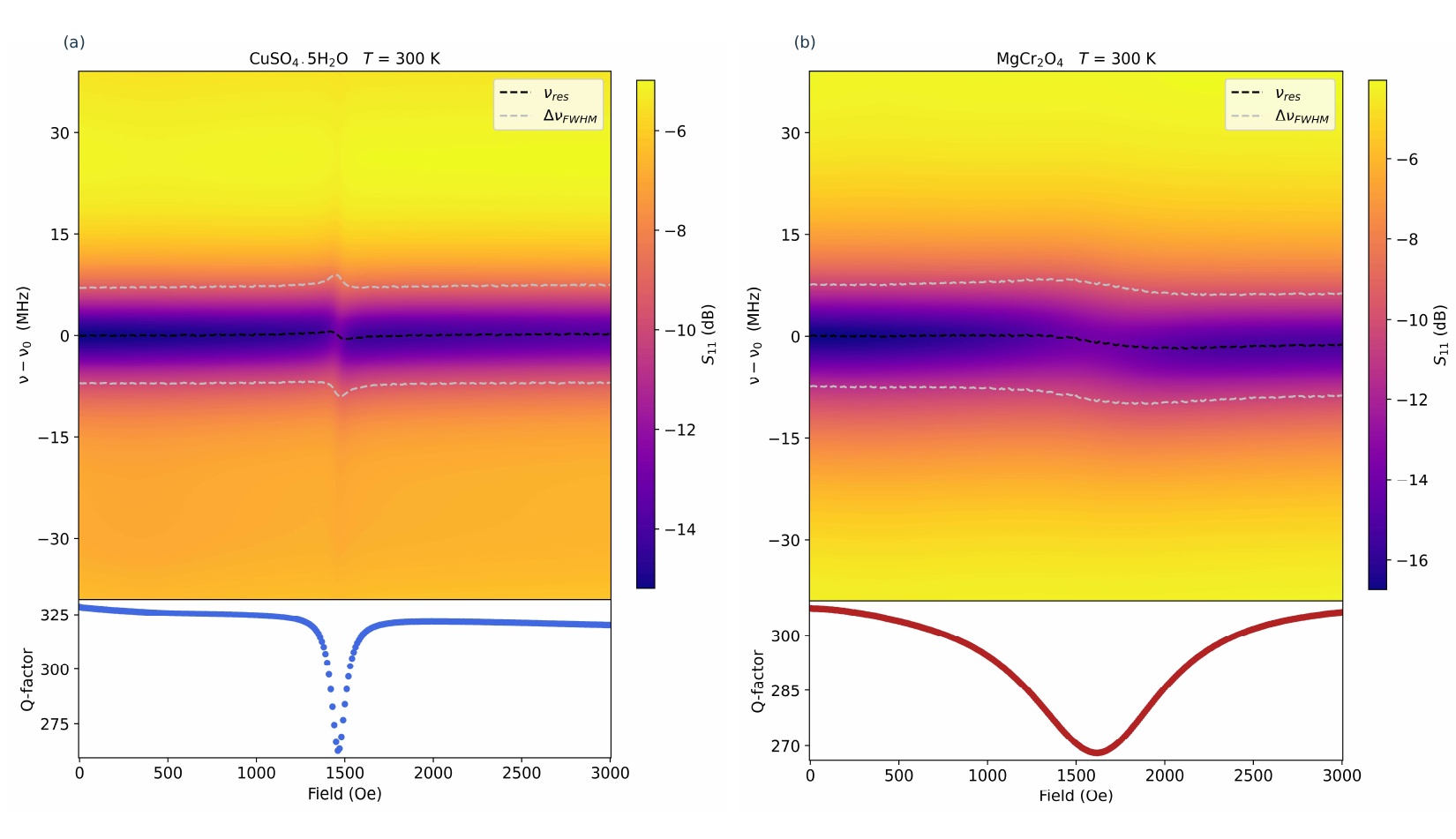}
        \caption{(a) Diagnostic plot for continuous wave ESR spectra for \cusulf\ at room temperature using set-up A with $\nu_{0}=4.64$~GHz. The top section shows the color-plot of the magnetic field, microwave frequency, and scattering parameter $\mathrm{S_{11}}$, created by taking a VNA trace every 10 Oe step of the orthogonal DC field ($B_0$). Each VNA trace was then fit to a standard Lorentzian to obtain the full width at half minimum (FWHM) and thus the $Q$-factor, $Q=\nu_{\rm res}/\Delta\nu$. The black dotted line marks the location of the minimum ${S_{11}}$, characterizing the resonance frequency, while the white dotted line marks the edges of the FWHM for each trace. The bottom section shows the fitted $Q$ factors as a function of field, allowing for our results to be compared to traditional ESR plots. The slight asymmetry and drift were determined to be caused by the real component of susceptibility ($\chi'$) changing as a function of field and is best explained by the high filling factor making the experiment more sensitive to these slight resonant frequency changes, as previously described in Ref.~\cite{yang_investigation_2023}. (b) Same plot for \magcro\, with $\nu_{0}=4.61$~GHz.} 
        \label{fig:ResultSetupA}
    \end{figure*}

\subsection{Sample-Resonator Mounting}

The sample-resonator union in setup A was mounted onto a 3D-printed holder with an open rectangular slot to firmly secure the bottom section of the resonator away from the sample loop and return flux paths. This slot was then mounted onto an electric transport puck built for the PPMS multi-function probe (MFP) so that the entire setup was attached to the rest of the probe. The bottom of the 3D printed slot was adhered with super glue to ensure the setup would not move during the experiment. The next step was to feed through a semi-rigid coax cable, threading it through the center hole on the head section of the MFP. One of the more tedious parts of the design was attaching the coupling loop to the bottom of this coax line. For the room temperature setup described here (setup A), the coupling loop was made of a stripped piece of coax where the central conductor was twisted and soldered to the outer conductor to form an inductive coupling loop. This coax termination transmitted the VNA signal into the loop-gap resonator, which then interacted with the sample. A combination of Teflon and adhesive was used to secure the coupling loop to the side of the MFP probe sample space, whose orientation is shown in Fig.~\ref{fig:FirstFig}(e). Stabilizing the coupling loop was a crucial part of taking these first room temperature measurements, as any movement would produce an unwanted signal superimposed on top of the material spectra. It should be noted that while the varnish method of stabilizing the resonators and coupling was useful, future design will utilize a more standard designed structure with brass screws and mounting components. 

\subsection{Measurement Protocol}
 
The DC magnetic field of the PPMS was applied along the vertical direction and was then swept between $B_0\!=\!0$~Oe to $3000$~Oe in steps of 10~Oe. At each point, the field was allowed to equalize for a few seconds before a ${S_{11}}$ trace was taken, where ${S_{11}}=P_{\rm reflected}/P_{\rm incident}$ is the RF power reflection coefficient at a given frequency. The raw data consisted of csv files of the VNA traces taken over a 1.5 GHz span centered on each sample-cavity resonance frequency. To get an overview of the data and observe if hybridization of the cavity polaritons and material spins occurred, the ESR spectra would be visualized as a 3D colormap by plotting the ${S_{11}}$ vs. frequency (GHz) traces as a function of $B_0$. This technique follows those used by other studies \cite{ball_loop-gap_2018,everts_ultrastrong_2020, Libersky_Design2019}.

\section{Results}

\subsection{Room Temperature}

The room-temperature results (``diagnostic plots'' using setup A) for both \cusulf\ and \magcro\ in their uncorrelated paramagnetic regime are presented in Fig.~\ref{fig:ResultSetupA}(a) and Fig.~\ref{fig:ResultSetupA}(b), respectively. Both compounds show a clear ESR response at room temperature, as visible, for \cusulf, from the broadening by around $4$~MHz of the VNA frequency lineshape $\Delta\nu$ beyond the baseline full-width at half-maximum of $\Delta\nu_0\approx14$~MHz and slight shift of the peak frequency $\nu_{\rm res}$ with respect to $\nu_{0}$. For \magcro, broadening by around $2$~MHz occurred from the $\Delta\nu_0\approx15$~MHz baseline. Calculating the $Q$-factor $Q(B_0) = \nu_{\rm res}({B_0})/\Delta\nu$ and plotting versus the applied magnetic field yields an unaligned $g$-factor of $g\approx 2.27$ for \cusulf, while the line-width of the ESR response was measured to be $\Delta B =90$~Oe full-width-half maximum, centered at a resonance field of $B_{\rm res}= 1460$~Oe. \magcro, on the other hand, displayed a much broader response, characteristic of its highly correlated paramagnetic spins, with a Weiss constant of $\Theta_{\rm W}= -346$ K for \magcro~\cite{bai_magnetic_2019} compared to $\Theta_{\rm W}= -0.6$ K for \cusulf. The measured $g$-factor for \magcro is $g\approx 2.03$, while the line-width was $\Delta B = 850$~Oe FWHM, centered at $B_{\rm res}=1620$~Oe. Both $g$-factor measurements agree with previously reported results \cite{blanchard_MgCr2O4_1986, bissey_differentiated_1995}. The cavity's quality factor (measured off-resonance) was approximately 320 in these conditions. While this low-quality factor can hinder sensitivity, the large filling factor ($\eta > 0.4$) can compensate for this deficiency, allowing us to reach the regime of strong spin-photon hybridization at low temperatures. 

\subsection{Data Analysis}

\begin{figure*}[t]
        \centering
    \includegraphics[width=1\textwidth]{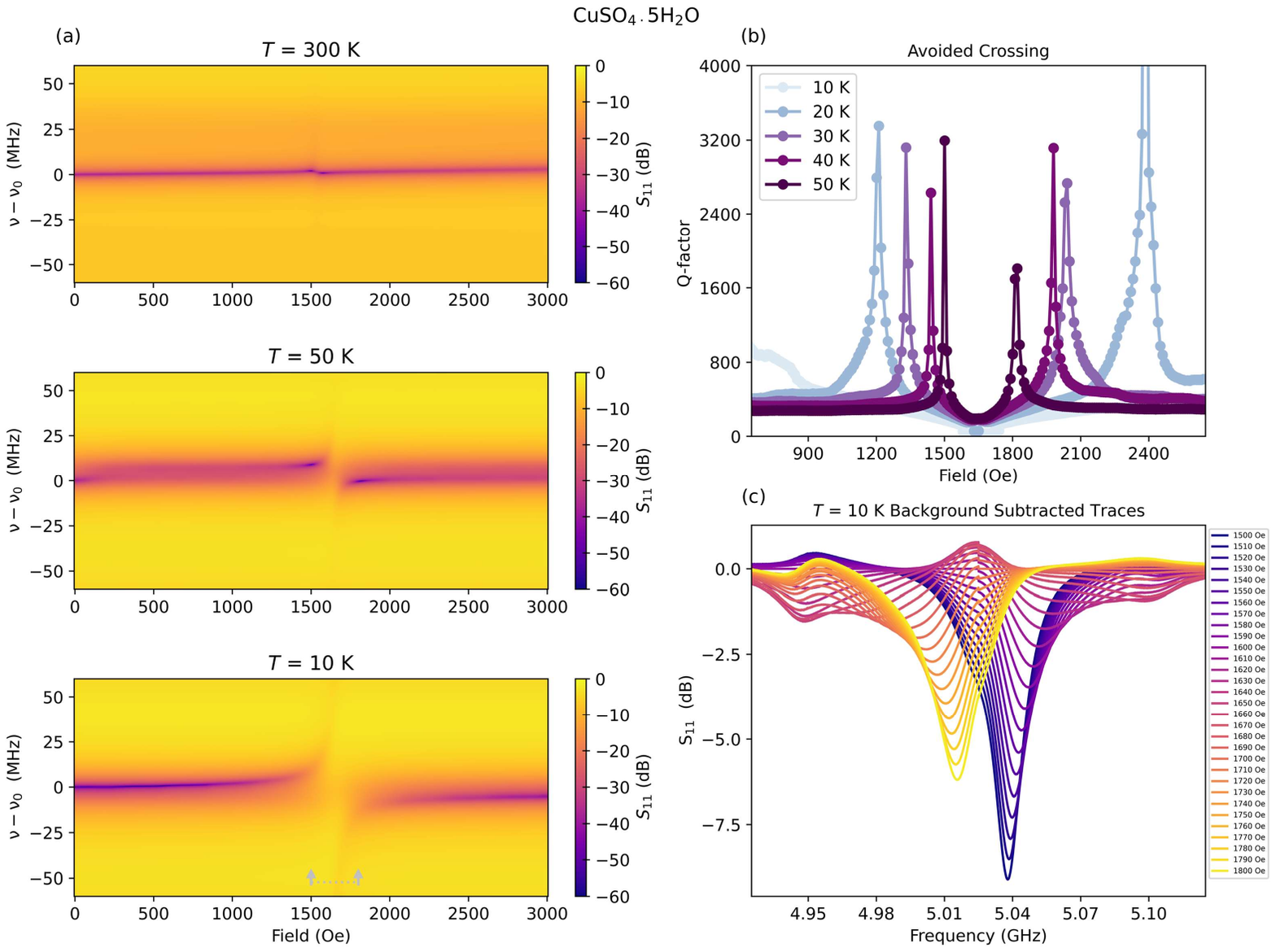}
        \caption{(a) Color plots for \cusulf\ temperature-dependent spectra  measured with setup B and displaying the increase in spin-cavity hybridization with decreasing temperature for $T$ = 300 K, 50 K, and 10 K. Each hybridized trace is centered around a $g$-value similar to the one reported with setup A, with the only deviations occurring for slight changes in crystal orientation between experiments. The data was collected during a single cool-down. Color plots are created from the raw $\mathrm{S_{11}}$ trace data at field increments of 10 Oe, and their analysis is described in the text and presented in the other sections of this caption. The frequency ($\nu_0=5.03$~GHz) is the resonance frequency of the sample-cavity system at zero field for each measurement. (b) $Q$-factor vs. magnetic field plots for 10 to 50 K in 10 K steps. (c)  Background subtracted VNA traces for the data between the grey arrows in the last plot of (a). Despite slight, unwanted features arising from an imperfect background subtraction, such as the positive regions in $S_{11}$, hybridized modes clearly form on either side of the central cavity-resonance frequency as the field sweeps towards and away from the expected peak ESR absorption. The legend to the side labels the field of each trace.}
        \label{fig:CuSO4setupB}
    \end{figure*}

\begin{figure*}[t]
    \centering
    \includegraphics[width=1\textwidth]{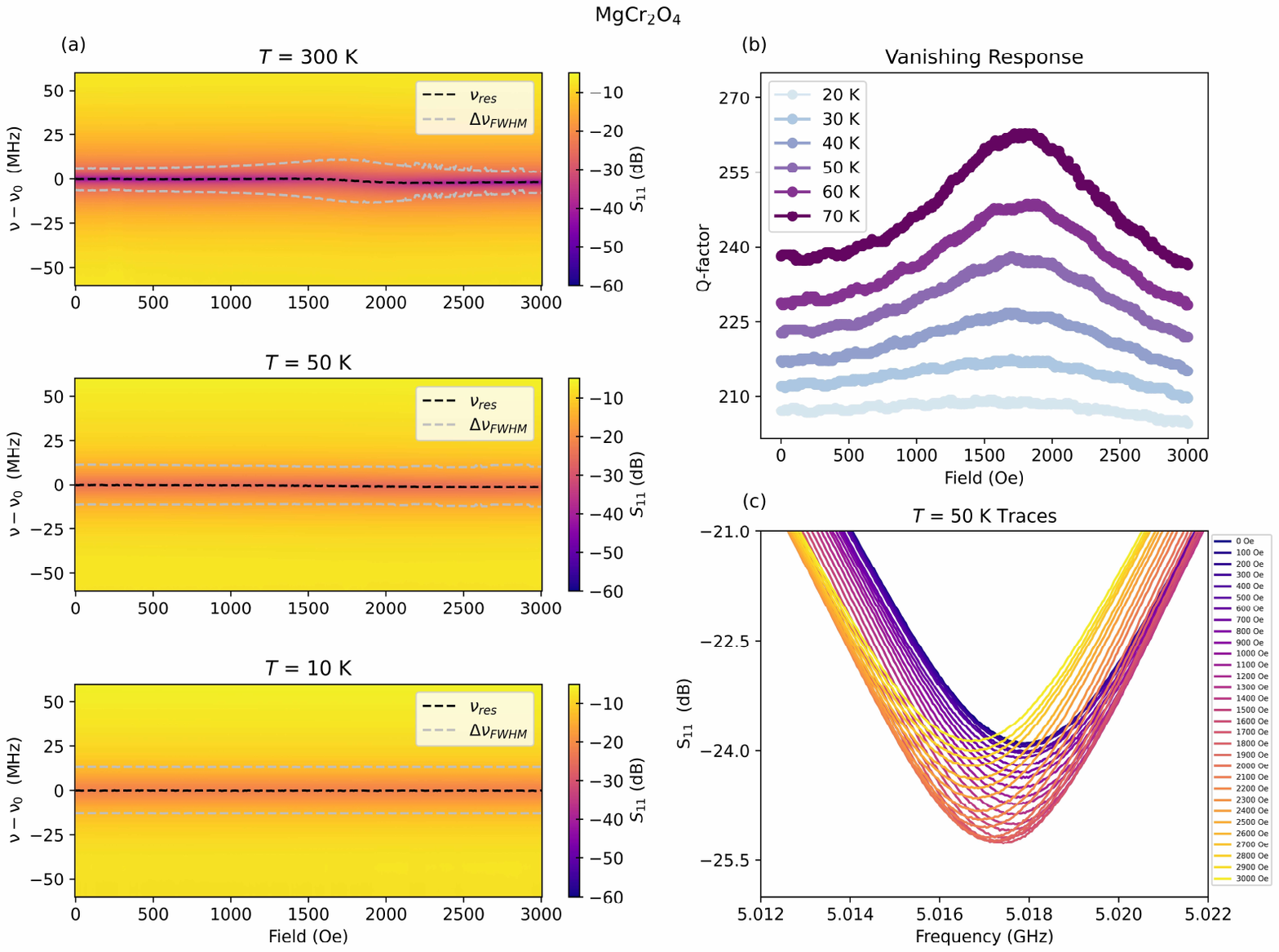}
        \caption{(a) Color plots for \magcro\ temperature-dependent spectra over the same three temperatures of $T$ = 300 K, 50 K, and 10 K as in Fig.~\ref{fig:CuSO4setupB}(a); however, no hybridization is observed as the temperature is decreased ($\nu_0=5.02$~GHz for this set-up). The $g$-factor is still consistent with the one reported earlier. Results were collected during a single cool-down from 300 K to 10 K in setup B as previously described in figure \ref{fig:CuSO4setupB}. (b) $Q$-factor vs. field plots for 20 to 70 K in 10 K steps. (c)  Raw VNA traces for \magcro\ over the entire swept spectra in 100 Oe steps for the 50 K measurement. While no observed response can be detected on the last two heat maps of (a), looking at the raw data here does show that both the sample-cavity resonance frequency and the linewidth change as we pass the resonant matching condition in a similar fashion as seen for \cusulf.  The side legend labels the field of each trace.}
        \label{fig:MgCr2O4setupB}
    \end{figure*}

Results obtained at low temperatures using setup B are presented for \cusulf\ and \magcro\ in Fig.~\ref{fig:CuSO4setupB} and \ref{fig:MgCr2O4setupB}, respectively. Before discussing these results, we present our data analysis approach in more detail, given that several strategies to reduce the raw data from the VNA are possible once the cavity and sample excitation mode hybridize. 

One of the key advantages of the VNA method of ESR detection is being able to separate both the real ($\chi'$) and imaginary ($\chi"$) components of complex susceptibility simultaneously~\cite{Libersky_Design2019,yang2023_CMP}. In conventional ESR, only the reactive ($\chi'$) or dispersive ($\chi"$) component is measured on a single resonance frequency. This can be directly detected via the reflected microwave power, or the signal can commonly be converted to a DC voltage via a Schottky diode, and is related to spectrometer parameters of operation through~\cite{rinard_loop-gap_2005} $$V_{\rm ESR}= \chi" \eta Q_L \sqrt{Z_0 P},$$ where $Z_0$ is the characteristic impedance of the transmission line (typically $50\Omega$), $P$ the input microwave power to the sample-resonator system, and $Q_L$ is the loaded quality factor defined as $Q_L =\nu_{\rm res}/\Delta \nu $ with the sample inside the resonator. Many commercial ESR spectrometers utilize an automated frequency control (AFC) circuit to stabilize the cavity resonance and optimize the reflected microwave output for a specific sample size \cite{poole_electron_1982}. It is possible to approximate this frequency stabilized resonance and thus observe the data in a traditional ESR format by tracking how $Q_L$ changes with DC magnetic field, similar to the techniques of other studies \cite{Collet2018,joshi_adjustable_2020}. There exist many analysis techniques originating from the superconducting circuits community to accurately determine the quality factor of a resonating circuit from a VNA trace, whether it be a Smith or $|S_{11}|$ plot, and reduce the effects of background signal from environmental factors \cite{petersan_1998_Q}.


\subsection{Cryogenic Response and Hybridization}

It becomes difficult to accurately fit a VNA trace when the cavity-spin system becomes hybridized. This situation occurs in setup B when \cusulf\ is cooled below $T=50$~K, see Fig.~\ref{fig:CuSO4setupB}(a) which displays a strong avoided crossing phenomena between the cavity mode and the Zeeman split paramagnetic excitation of the sample reaching up to $\Delta B \approx 1200$~Oe. In that situation, the original central resonance frequency disappears, and two weaker hybridized modes emerge, which become difficult to fit and determine a full width at half max (FWHM) value to extract the $Q$-factor. Compounding this effect is the periodic background signal coming from the length of coax cable used in the experiment. For this reason, the $Q$-factors for each trace, shown in Fig.~\ref{fig:CuSO4setupB}(b) are determined from a simplified fit that extracts the maximum value of $\mathrm{|S_{11}|}$ at the resonance position and collects the nearest data points at half that value to determine the frequency FWHM and the $Q$-factor. This procedure neglects the raised $\mathrm{|S_{11}|}$ baseline produced by the background coax-signal, resulting in a broader FWHM determination than the actual value. However, the negative impact is only observed at the resonance matching condition when the signal broadens to its most considerable extent for \cusulf\ and results in the $Q$-factor appearing to reach a zero value. This quick drop to $Q\approx0$ at resonance matching can be seen for the two lowest temperatures in Fig.~\ref{fig:CuSO4setupB}(b).  While plotting $Q(B)$ is primarily used in traditional ESR, here it allows a depiction of how spin-polarition coupling strength is increasing, as our zeroth-order Lorentzian fitting program outlined in the text can accurately obtain the VNA dip linewidth up to the regime where hybridization makes fitting with the coax background impossible and $Q \rightarrow 0$ (as seen in the plot). 

In contrast to \cusulf, the temperature dependent results for \magcro, Fig~\ref{fig:MgCr2O4setupB}(a), display no strong hybridization. The $Q$-factor vs. magnetic field plots shown in Fig~\ref{fig:MgCr2O4setupB}(b) allow for a traditional ESR interpretation, and the decreasing paramagnetic response as the sample approaches $T_N\approx 15 \: K$ is consistent with previous reports for \magcro \cite{blanchard_MgCr2O4_1986}.

Usually, matching at the resonance condition between the DC field and the microwave frequency produces a decreased loaded quality factor that can be measured as the ESR response. This behavior was observed in our set-up A, see Fig.~\ref{fig:ResultSetupA}. This result is usually explained in textbooks \cite{poole_electron_1982} (and observed in experiments \cite{Collet2018}) as more energy is dissipated by flipping paramagnetic spins or generating magnons, resulting in a quality factor decrease. However, this is not always the case, as can be seen in Fig.~\ref{fig:MgCr2O4setupB}(b), where the quality factor increases upon reaching the resonance condition. Instead, this data can be better understood as the magnetic material changing the reactance and thus the coupling of the cavity-sample system to the transmission line. Whether or not the magnetic material causes the quality factor to increase or decrease depends on how close the off-resonance Lorentzian is to critical coupling and if the resonance frequency change associated with matching takes the peak closer or further to the critical coupling condition. If the system is critically coupled or close to it, reaching resonance will always decrease the quality factor. Still, an increasing quality factor can be observed when the setup is far enough from critical coupling. This shows why it is essential to properly characterize the state of coupling for a microwave setup. The magnitude and positioning of the ESR response is still sample-dependent, so even off of the critical coupling condition, we are able to obtain significant results. Once the stub tuner was adjusted to reach closer to critical coupling, the same temperature-dependent results in Fig.~\ref{fig:MgCr2O4setupB}(b) were repeated, but this time with a quality factor that decreased as expected (results not show). 

In Fig.~\ref{fig:CuSO4setupB}(c) shows every VNA trace taken between 1500 to 1800 Oe for \cusulf, where strong hybridization is seen to occur between the two grey arrows located in the bottom plot of Fig.~\ref{fig:CuSO4setupB}(a).

These traces have each been modified by subtracting the same background trace from each profile. This is to remove the repeated coax signal previously mentioned and observe the pure cavity-sample response of each. However, it can be seen that periodic irregularities still exist in that there are slight protrusions above 0 dB at the frequencies of 4.950~GHz, 5.025~GHz, and 5.100~GHz. It should be noted that it was confirmed that this background signal was from the coax cable by using the well-known expression: 
$$\mathrm{Cable\:Length} = \frac{c}{2\Delta f \sqrt{\epsilon_R}},$$ which produces a length of 1.4 m for Teflon's dielectric constant, and a frequency dip separation of 75 MHz. This matched the length of the cable used to within a centimeter, confirming the signal's origin. The background trace was created by stitching together the trace at 1600 Oe from the lowest frequency collected to the central resonant frequency  ($\nu_{\rm res}= 5.025$~GHz) with the trace at 1700 Oe from the central resonant frequency to the highest frequency collected. 
This allows the background signal to be reconstructed by adding opposing trace sections near the avoided level crossing that reveal opposite sections of the true background that the sample-LGR resonant response is superimposed onto. For VNA traces of \magcro\ over its entire swept spectra in 100 Oe intervals, shown in Fig.~\ref{fig:MgCr2O4setupB}(c), no background trace is subtracted since no strong hybridization occurs, meaning the frequency region of interest does not span a large enough region for the coax signal to obscure any results. With the background trace subtracted for \cusulf, the faint hybridized modes are seen for traces around the center of $B=1660$~Oe.

    \begin{figure}[t]
    \centering
    \includegraphics[width=1\textwidth]{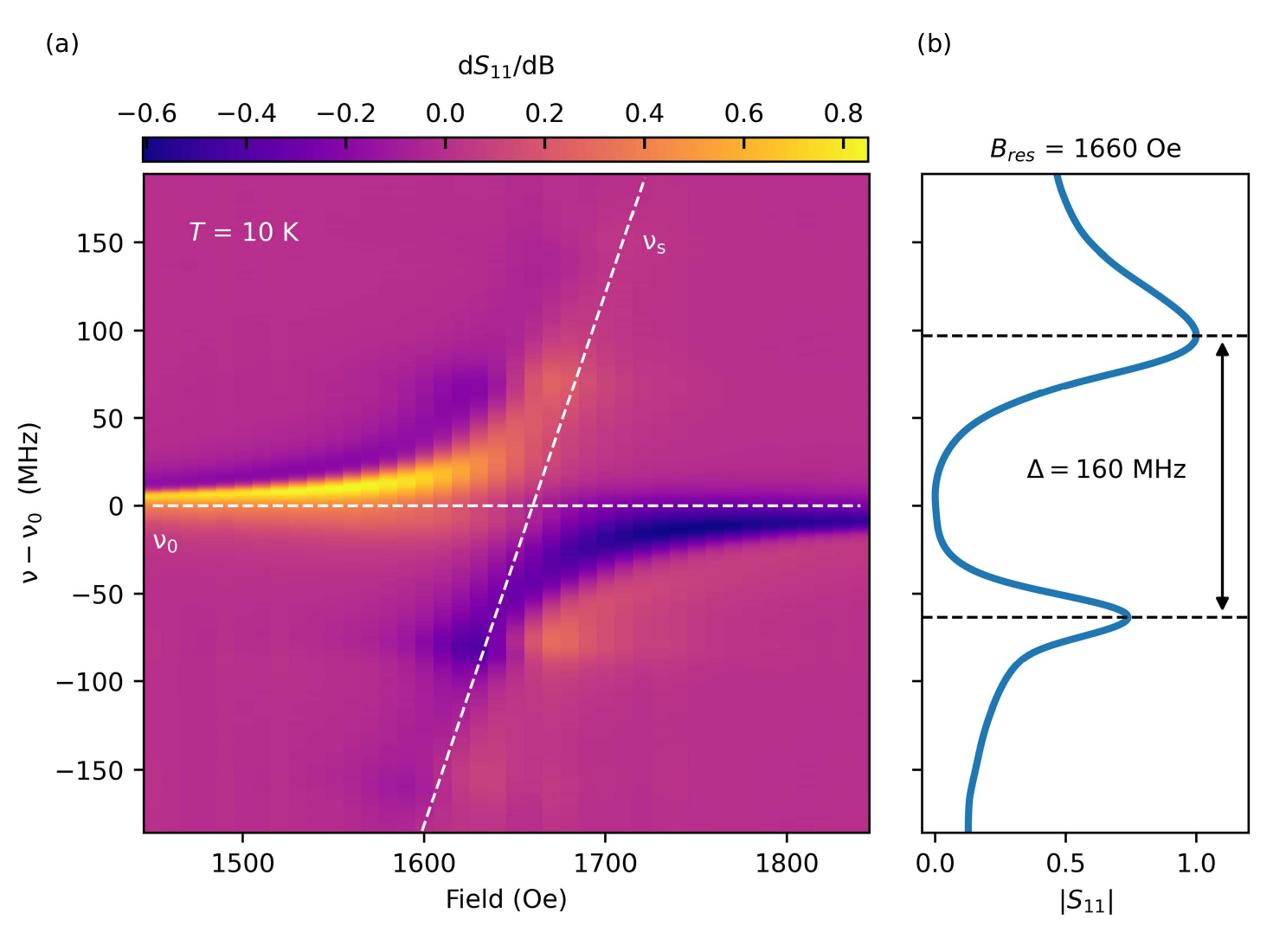}
        \caption{(a) Derivative of $S_{11}$ with respect to magnetic field ($B$) as a function of frequency and field for \cusulf\ at 10 K. While this merely plots the field derivative of the data shown in the bottom plot of Fig.~\ref{fig:CuSO4setupB}(a), taking the derivative in such a way allows for the removal of the majority of unwanted resonances caused by the coax cable. However, the effects of such cable resonances can still be seen in the amplitude increases at $\pm 75$~MHz, where they intersect with the polariton modes. The horizontal white dashed line ($\nu_0$) marks the primary cavity mode, while the diagonal white dashed line marks the \cusulf\ eigenfrequency ($\nu_s$) that intersects with the cavity mode. (b) Reconstructed and normalized $|S_{11}|$ as a function of VNA frequency at $B=1660$~Oe. By taking the integral of the $S_{11}$ derivative at this resonance field, the vacuum Rabi splitting ($\Delta$), and thus the magnon-polariton coupling strength ($G$), can be approximated as $\Delta = 2G/(2\pi) \approx 160$~MHz. Asymmetry in the two Lorentzian peaks is likely caused by a combination of unwanted interference with the upper cable resonance and a slight drift in the bare cavity resonance during the measurement. For this reason, a conservative range of our coupling strength is between $G/2\pi \approx 50$ to $80$~MHz. }
        \label{fig:S11_Derivative}
    \end{figure}

\subsection{Interpretation}

Overall, the primary difference between the two sample spectra studied down to $T=10$~K is that \cusulf\ is seen to strongly hybridize with the cavity mode, increasing the degree of hybridization with a decreasing temperature. At the same time, \magcro\ shows no hybridization with the resonator environment despite experimental considerations such as filling factor, resonance frequency, and mode-volume confinement being the same.

An expression showing the coupling strength between antiferromagnetic spins and the microwave cavity is provided by \cite{everts_ultrastrong_2020}: 
$$G=\sqrt{\frac{\pi N \nu_{\rm res} \mu_0}{\hbar V_{\rm res}}}\mu_B \eta,$$
where $N$ is the number of magnetic ions in the sample, $V_{res}$ is the cavity mode volume, and $\eta$ is the filling factor. This expression illustrates the dependence of coupling strength on the ratio of cavity resonance frequency ($\nu_{res}$) to mode volume characteristic of microwave resonators and explains the necessity of small-volume LGRs. To approximate the coupling strength for \cusulf\ at $T=10$~K, the 1st derivative of $S_{11}$ with respect to the field ($B$) is taken and then plotted in a fashion similar to \cite{Yu_2023}. The results are shown in Fig.~\ref{fig:S11_Derivative}(a). Taking the derivative of $S_{11}$ in this way helps to remove the constant signal created by the coax cable and allows a more precise determination of the avoided level crossing. With a large part of the coax signal removed in the derivative, numerically integrating the derivative trace at $B_{res}\approx 1660$~Oe gives a decent approximation of the vacuum Rabi splitting, as shown in Fig.~\ref{fig:S11_Derivative}(b). The gap between peaks is given as $\Delta = 2G/(2\pi)\approx 160$~MHz. The asymmetry in the peaks is likely due to unwanted features from the coax signal that still influence the reconstructed $S_{11}$ data. Based on this figure, the coupling strength is estimated to be $G/2\pi \approx 50$ to $80$~MHz, with a more precise estimate requiring experimental improvements to eliminate the cable resonances or modeling the spin-cavity Hamiltonian. 

The lack of hybridization for \magcro\ is speculated to arise from its more strongly correlated paramagnetic spins that produce a broad ESR response even at room temperature. However, observing sample-cavity coupling differences between these two samples demonstrates that our setup probes both the hybridization and magnetic response across a wide range of temperatures in a relatively simple experimental architecture.

\section{Conclusion}

This manuscript reports the design and implementation of loop-gap resonators (LGRs) for electron spin resonance (ESR) on quantum materials and hybrid quantum system studies using  microwave polariton modes. The experimental setup is low-cost and implemented in cryogenic equipment commonly found in laboratories worldwide. 

First experiments were conducted on single-crystals of paramagnetic salts down to $T=10$~K. Although the precise interpretation in terms of microscopic models is still under development, our findings indicate that sample-embedded loop gap resonators (LGRs) with high filling factors serve as unique and flexible probes for studying magnetic susceptibilities and strong spin-polariton coupling in hybrid quantum systems. Expanding upon previous studies that validate the viability of LGRs \cite{ball_loop-gap_2018, everts_ultrastrong_2020,Libersky_theory_magnon}, our work applies them to investigate spin-liquid candidate materials of interest in quantum materials research. We highlight that this technique bridges the gap between well-established, traditional magnetic resonance studies and novel investigations of hybrid-quantum systems and heterostructures at low temperatures.

Our report hopes to assist other experimental groups in developing effective microwave resonance methods for studying bulk magnetic samples. Our method not only facilitates the acquisition of magnetic susceptibility data but also enables the accessible modification of the spin's electromagnetic environment by inducing strong spin-polariton modes. By implementing our two setups on a widespread controllable platform, the Physical Property Measurement System (PPMS), along with accessible microwave equipment (stub tuner and network analyzer), and by providing a detailed step-by-step analysis process, including aspects not commonly reported in the literature, we contribute to the broader understanding of these techniques.

In our future research endeavors, we plan to enhance our experimental setup by introducing precision-machined resonators, optimizing the microwave bridge, and conducting experiments at lower temperatures in a dilution refrigerator. This upgraded configuration goes beyond mere integration with a network analyzer; it promises heightened control and versatility. Additionally, our upcoming studies aim to explore the time-domain response of the strongly coupled states identified in this work. We expect that these advancements will not only contribute to our understanding of an effective and accessible technique but also drive progress in the fields of condensed matter and quantum information.


\acknowledgments
The work at Georgia Tech was supported by the National Science Foundation under award NSF-DMR-1750186 (A.K.J., M.M.). This work was performed, in part, at the Center for Integrated Nanotechnologies, an Office of Science User Facility operated for the U.S. Department of Energy (DOE) Office of Science. Sandia National Laboratories is a multimission laboratory managed and operated by National Technology and Engineering Solutions of Sandia, LLC, a wholly owned subsidiary of Honeywell International, Inc., for the U.S. DOE’s National Nuclear Security Administration under contract DE-NA-0003525. The views expressed in the article do not necessarily represent the views of the U.S. DOE or the United States Government. We are grateful to Oleg Starykh for nudging us into the realm of ESR in quantum magnets and Henrik R\o{}nnow for pointing out invaluable references to us.


\providecommand{\newblock}{}

\end{document}